\documentclass[11pt]{article}
\pdfoutput=1
\usepackage{jheppub}
\usepackage{graphicx,float,flowchart}
\usepackage{amssymb}
\usepackage{subcaption}
\usepackage{amsmath}
\usepackage{slashed}
\usepackage{hyperref}
\usepackage{caption}
\usepackage{xcolor}
\usepackage{dsfont}
\usepackage{verbatim}
\usepackage{setspace}
\usepackage{slashed}
\usepackage{xcolor}
\usepackage{physics}
\usepackage{ragged2e}

\newcommand\beq{\begin{equation}}
\newcommand\eeq{\end{equation}}

\def\shrug{\texttt{\raisebox{0.75em}{\char`\_}\char`\\\char`\_\kern-0.5ex(\kern-0.25ex\raisebox{0.25ex}{\rotatebox{45}{\raisebox{-.75ex}"\kern-1.5ex\rotatebox{-90})}}\kern-0.5ex)\kern-0.5ex\char`\_/\raisebox{0.75em}{\char`\_}}}

\title{Non-local Entanglement and Fast Scrambling in De-Sitter Holography}
\author{Hao Geng}
\affiliation{Department of Physics, University of Washington, Seattle, WA, 98195-1560, USA}
\emailAdd{hg666@uw.edu}
\preprint{\today}
\abstract{We study holographic entanglement and information scrambling in de-Sitter (dS) space in the context of the DS/dS correspondence. We find that our previously identified non-local entanglement structure of dS vacua can be extended out of the time-reflection symmetric slice. We extend the geometry to a two-sided configuration and calculate the zero-time mutual information between two intervals on different sides when there is a localized shock wave in the bulk. Interestingly, we find that the information scrambling time saturates the fast scrambler bound proposed by Sekino and Susskind and that the shock wave renders a wormhole to be traversable. Furthermore, we calculate a two-sided out-of-time-ordered correlator (OTOC) in the late time regime and we see that, before scrambling, it exponentially grows with an exponent whose value saturates the maximal bound of chaos proposed by Maldacena, Shenker and Stanford. At the end, we provide an explanation why the exponential growing of the late-time OTOC with the maximal bound of chaos saturated and the traversability of the wormhole are simple results of the non-local entanglement structure and point out that this is a realization of the ER=EPR proposal. }
\begin{document}
\maketitle
\section{Introduction}
Equipped with the AdS/CFT correspondence \cite{Maldacena:1997re,Gubser:1998bc,Witten:1998qj,Aharony:1999ti}, people have understood a lot of fascinating aspects of quantum gravity in the last two decades. These include the information theoretic nature of some aspects of quantum gravity \cite{Harlow:2014yka,Harlow:2018fse,Ryu:2006bv,Ryu:2006ef,Hubeny:2007xt} and the chaotic nature of the dynamics of black holes \cite{Shenker:2013pqa,Roberts:2014isa,Maldacena:2015waa,Jensen:2016pah} in anti-de-Sitter (AdS) space. However, due to the lack of a well-established holographic framework of quantum gravity in de-Sitter (dS) space, the parallel studies of quantum gravity or holography in dS space is very limited. Recently, motivated by several lessons we learned from studies in AdS and a general formulation of holography, known as the surface/state correspondence \cite{Miyaji:2015yva}, the author studied several information theoretic aspects of dS holography in \cite{Geng:2019ruz,Geng:2019bnn} using the DS/dS correspondence\footnote{It deserves to be mentioned that there is another holographic proposal for dS quantum gravity- the dS/CFT correspondence\cite{Strominger:2001pn,Maldacena:2002vr}. Recent studies in that direction include \cite{Sato:2015tta,Fernandes:2019ige,Narayan:2015vda,Narayan:2017xca,Narayan:2019pjl,Narayan:2020nsc,Arias:2019pzy,Arias:2019zug,Yokoyama:2020tgs,Narain:2018wlz}. Other interesting studies of dS entanglement using the direct approach (a quantum field probe) includes \cite{Akhtar:2019qdn,Banerjee:2020ljo,Bhattacharya:2019zno,Bhattacharya:2018yhm}. }\cite{Alishahiha:2004md,Karch:2003em}. We discovered an unconventional non-local entanglement structure of the dual field theory system on the time-reflection symmetric slice which was also briefly mentioned by Miyaji and Takayanagi \cite{Miyaji:2015yva}. \footnote{It deserves to be mentioned that this discovery is consistent with the replica-trick calculations in \cite{Dong:2018cuv,Lewkowycz:2019xse} which shows that the density matrices for small subsystems are maximally mixed.}And we suggested that the field theory dual should be maximally chaotic.\footnote{Other recent and old works addressing de-Sitter chaos include \cite{Bhattacharyya:2020rpy,Anninos:2018svg,Nomura:2011dt,Susskind:2011ap,Liu:2019nit,Seo:2019wsh,Aalsma:2020aib}.} Extending our previous studies to understand some dynamical aspects of dS holography, analogous to those dynamical aspects in AdS discussed by Shenker and Stanford \cite{Shenker:2013pqa}, would be a very important step towards an appreciation of quantum gravity in our universe. And this is the aim of this work. We hope that these peculiar developments will encourage people to reconsider dS holography.

In this paper, we will put our suggestion of maximal chaos\footnote{Note that in this paper we will not distinguish between maximal chaos and fast scrambling of quantum systems. Even though, recent studies such as \cite{Xu:2019lhc} question this equivalence.} in dS holography on a solid footing by studying the information scrambling of the dual field theory dynamics and see that it is a fast scrambler \cite{Sekino:2008he}. 

We firstly extend the exact non-local entanglement structure beyond the time-reflection symmetric slice. Then we will study the information scrambling using the so-called two-sided mutual information in a shock wave geometry (as that by Shenker and Stanford in \cite{Shenker:2013pqa}) and see that the scrambling time indeed saturates the bound for a fast scrambler \cite{Sekino:2008he}. This tells us that dS space is indeed a maximally chaotic system or a fast scrambler. To do this, we will do Kruskal extension to extend the dS$_{D}$ to a two-sided geometry where there are two parts of the dual field theory system living on different sides. Interestingly, we notice that the shock wave renders the bulk wormhole to be traversable. After this, we will study the two-sided out-of-time-ordered correlator (OTOC) for a heavy bulk field and see that, before scrambling, the magnitude of the late-time growing exponent saturates the maximal chaos bound \cite{Maldacena:2015waa}  but it has a negative sign compared with the usual Lyapunov exponents (this is also noticed by the authors of \cite{Aalsma:2020aib} in a different context). Amazingly, we will see that the traversability of the wormhole and the negative sign of the late-time growing exponent are due to our exact non-local entanglement structure. This discovery tells us that we can use fast scramblers to do quantum teleportation as long as the information is encoded in a non-local entanglement structure. 

This structure of this paper is as following, in Sec.\ref{sec:basics} we will review different coordinate patches of de-Sitter space from the embedding space formulation, set up our notations through the paper and provide a lightening review of the DS/dS correspondence. In Sec.\ref{sec:ENES} we will firstly review the relevant parts of our previous works \cite{Geng:2019ruz,Geng:2019bnn}, the surface/state correspondence \cite{Miyaji:2015yva} and the exact non-local entanglement structure. Then we move on to extend this exact non-local entanglement structure to dS global patch and discuss the so-called cascade of dS quantum gravity. In Sec.\ref{sec:fastscra} we will study information scrambling in dS holography and show various interesting and subtle aspects.. At the end in Sec.\ref{sec:final} we conclude our paper with a summary and remark on possible future developments. 

\section{Basics of De-Siter Space and the DS/dS correspondence}\label{sec:basics}
In this section we set up our notations and discuss different parametrizations of de-Sitter space which will be frequently used in later sections. Then we give a lightening review of the DS/dS correspondence. Readers familiar with these basics can skip this section and come back when unfamiliar notations are encountered.

A D-dimensional De-Sitter space can be embedded into a D+1-dimensional Minkowski space as a hyperboloid\footnote{For the sake of convenience, we will set the length scale $L$ to be 1 in later discussions.} (See Fig.\ref{pic:hyperboloid}):
\begin{figure}
    \centering
   \includegraphics[width=10.1cm]{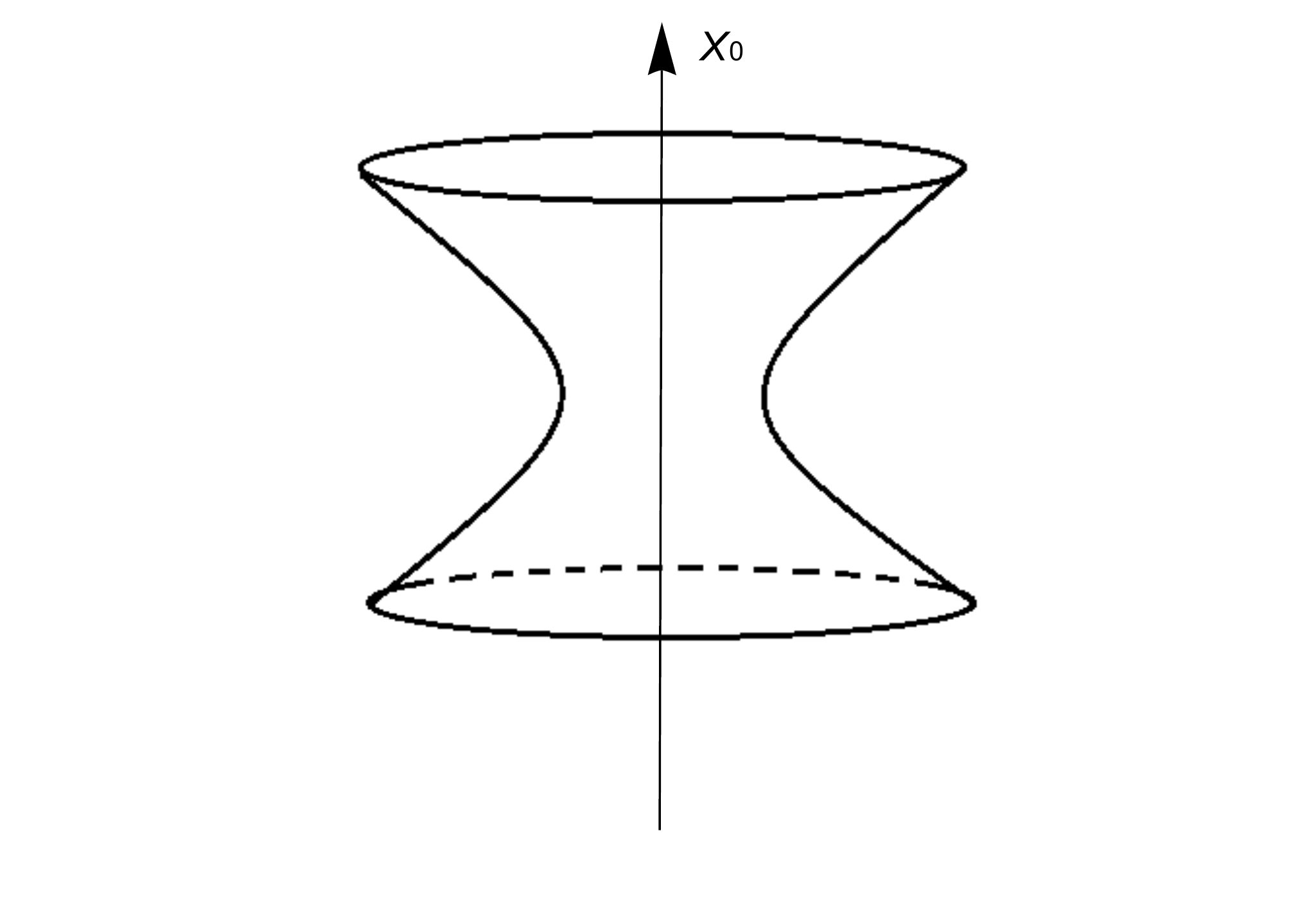}
        \caption{Embedding de-Sitter to a Higher Dimensional Minkowski. }\label{pic:hyperboloid}
\end{figure}
\beq
  -X_{0}^2+X_{1}^2+\cdots+X_{D}^2=L^{2}.\label{eq:def}
\eeq
Using this fact we can parametrize de-Sitter space in different ways and see different physics.
\subsection{Global Patch} 
\begin{figure}
    \centering
   \includegraphics[width=10.1cm]{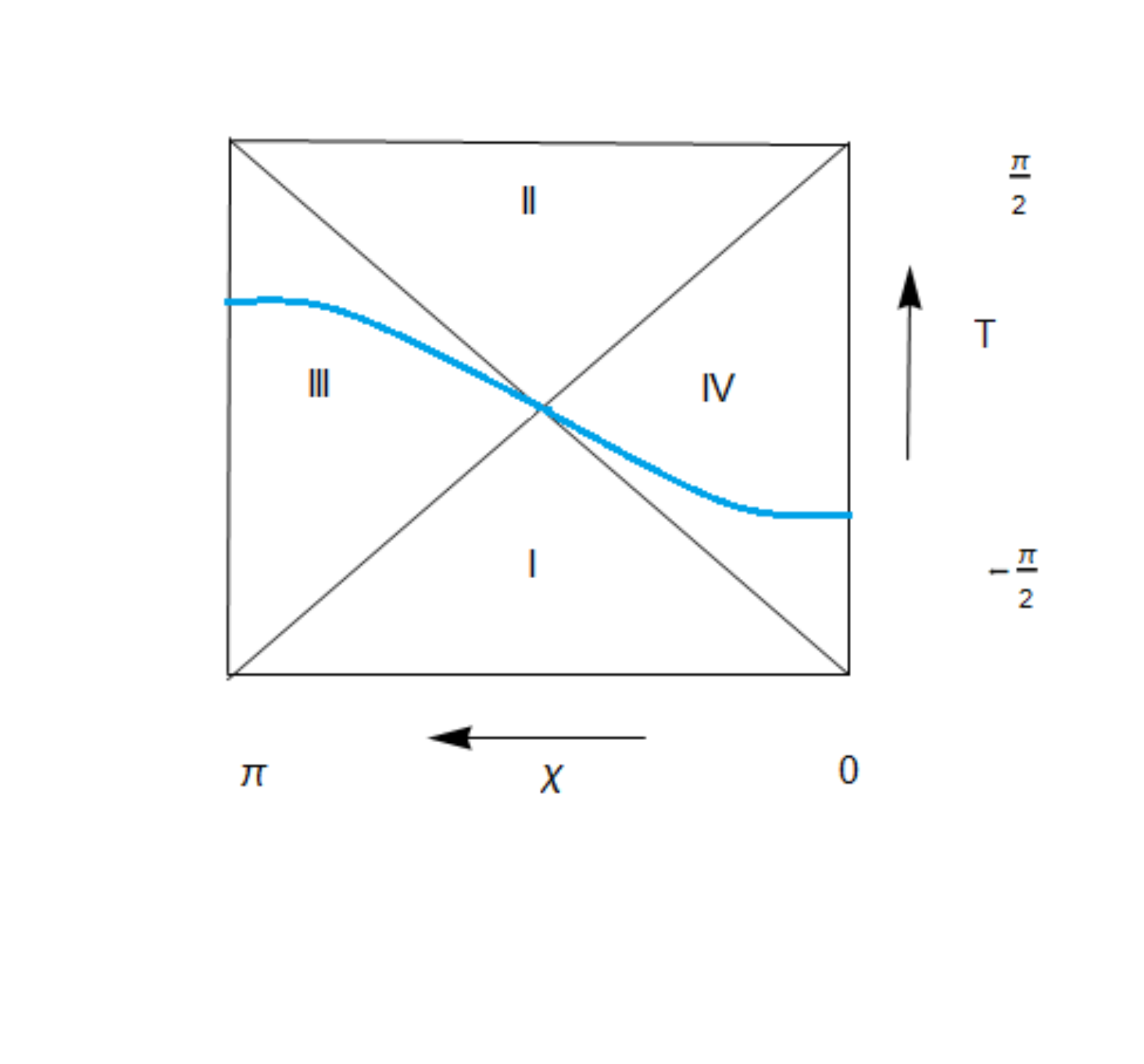}
        \caption{The Penrose diagram of de-Sitter. Global coordinate covers the whole square and the extended static coordinate only covers regions III and IV. The blue slice is a constant time slice in static coordinate. Each point on the diagram is a $D-2$ sphere.}\label{pic:penrose}
\end{figure}

 The embedding coordinates are parametrized as
\begin{equation}
  \begin{split}
X_{0}&=\sinh{\tau_{D}}\\
X_{i}&=\Omega_{i}\cosh{\tau_{D}}
  \end{split}   
\end{equation}
where $i=1,2,\cdots,D$, $\tau_{D}$ is called the ``DS global time'' and it covers the whole hyperboloid in Fig.\ref{pic:hyperboloid}. We have the following global metric
\beq
  ds^{2}=-d\tau_{D}^{2}+\cosh^{2}{\tau_{D}} d\Omega_{D-1}^{2}
\eeq
which is conformally equivalent to the following metric
\beq
\begin{split}
    T&=\arctan{(\sinh{\tau_{D}})}, T\in [-\frac{\pi}{2},\frac{\pi}{2}]\\
    d\tilde{s}^{2}&=-dT^{2}+d\chi^{2}+\sin^{2}\chi d\Omega_{D-2}^{2}.\label{eq:conformal}
\end{split}
\eeq
From this conformal equivalence we have the Penrose diagram for $D\geq3$ in Fig.\ref{pic:penrose}.

\subsection{Extended Static Patch}
The coordinates parametrization is as following
\beq
\begin{split}
    X_{0}&=\cos\beta_{D} \sinh t_{D}\\
    X_{1}&=\cos\beta_{D}\cosh t_{D}\\
    X_{j}&=\Omega_{j}\sin\beta_{D} 
\end{split}
\eeq
where $\beta_{D}\in[0,\pi]$, $j=2,3,\cdots,D$, $t_{D}$ is called the ''static time" and this coordinate covers only regions III and IV in Fig.\ref{pic:penrose}. The metric is given by
\beq
\begin{split}
  ds^{2}&=-\cos^{2}\beta_{D} dt_{D}^{2}+d\beta_{D}^{2}+\sin^{2}\beta_{D} d\Omega_{D-2}^{2}\\
  &=-\cos^{2}\beta_{D} dt_{D}^{2}+d\beta_{D}^{2}+\sin^{2}\beta_{D} (d\chi_{d}^{2}+\sin^{2}\chi_{d}d\Omega_{D-3}^{2})
  \end{split}
\eeq
so $t_{D}$ is the orbit of a time-like Killing vector field and the two diagonals on Fig.\ref{pic:penrose} are the horizons (where $g_{00}=0$).
\subsection{DS/dS$_{\text{global}}$ Patch}
\begin{figure}
    \centering
   \includegraphics[width=10.1cm]{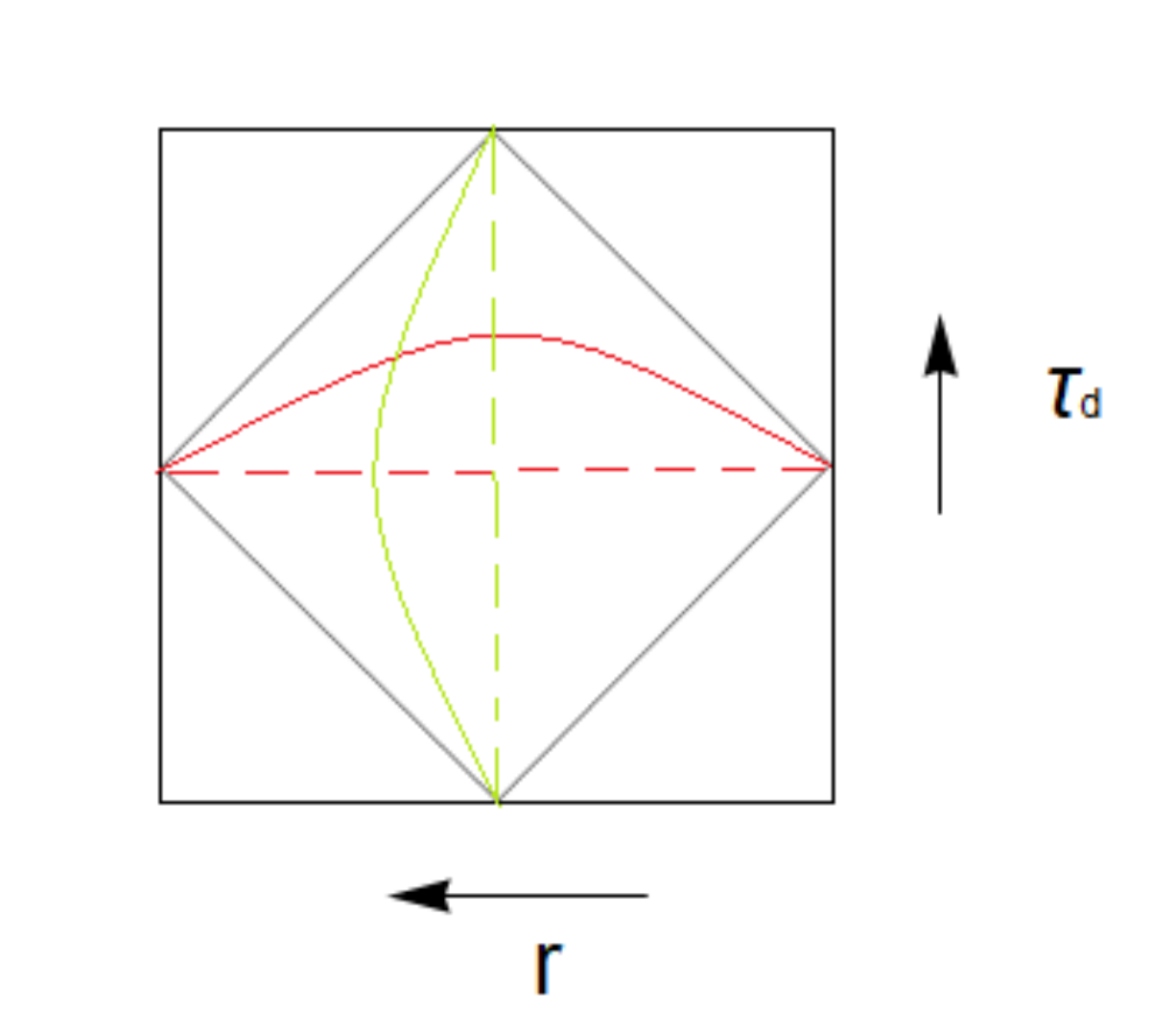}
        \caption{The Penrose Diagram of de-Sitter global patch sliced de-Sitter. The solid red slice is a constant $\tau_{d}$ slice and the dashed red slice is $\tau_{d}=0$ slice. The solid green slice is a constant $r$ (radial) slice and the dashed green slice is the $r=\frac{\pi}{2}$ (central) slice. Each point on the diagram is a (D-2)-sphere.}\label{pic:DS/dSglobal}
\end{figure}

This coordinate is based on the observation that we can slice a D-dimensional hyperboloid in Equ.\ref{eq:def} by (D-1)-dimensional hyperboloids and those (D-1)-dimensional hyperboloids, called dS slices, are covered by global coordinates:
\beq
\begin{split}
    X_{1}&=\cos{r}\\
    X_{0}&=\sin{r} \sinh{\tau_{d}}\\
    X_{j}&=\sin{r}\cosh{\tau_{d}}\Omega_{j}\label{equ:DS/dsglobal}
\end{split}
\eeq
where $r\in [0,\pi]$, $j=2,3,\cdots,D$, $\tau_{d}$ is called the ``dS global time" and this coordinate covers the central diamond in Fig.\ref{pic:DS/dSglobal}.
For the sake of convenience of later utility, we give the following coordinate transformation from dS-sliced coordinate to the global conformal coordinate Equ.\ref{eq:conformal}:
\beq
\begin{split}
    \sin r\sinh\tau_{d}&=\tan T\\
    \cos r&=\frac{\cos\chi}{\cos T}.\label{eq:transformation1}
\end{split}
\eeq
In this dS-sliced coordinate the metric is given by
\begin{equation}
    ds^{2}_{\text{(A)dS$_{D}$}}=dr^{2}+\text{sin(h)}^{2}(\frac{r}{L})\text{ }(-d\tau_{d}^{2}+\cosh^{2}(\frac{\tau_{d}}{L})d\Omega_{D-2}^{2})
\end{equation}
and it would reappear in Equ.\ref{equ:motivation} which motivates the DS/dS correspondence \cite{Alishahiha:2004md,Karch:2003em}. 

Here for later convenience in Sec.\ref{sec:ENES}, we blow up the $\Omega_{D-2}$ there to be foliated by $\Omega_{D-3}$:
\begin{equation}
   ds^{2}=dr^{2}+\sinh^{2}(r)\Big[-d\tau_{d}^{2}+\cosh^{2}(\tau_{d})(d\chi_{d}^{2}+\sin^{2}\chi_{d}d\Omega_{D-3}^{2})\Big].
\end{equation}
\subsection{DS/dS$_{\text{extended static}}$ Patch}
\begin{figure}
    \centering
   \includegraphics[width=10.1cm]{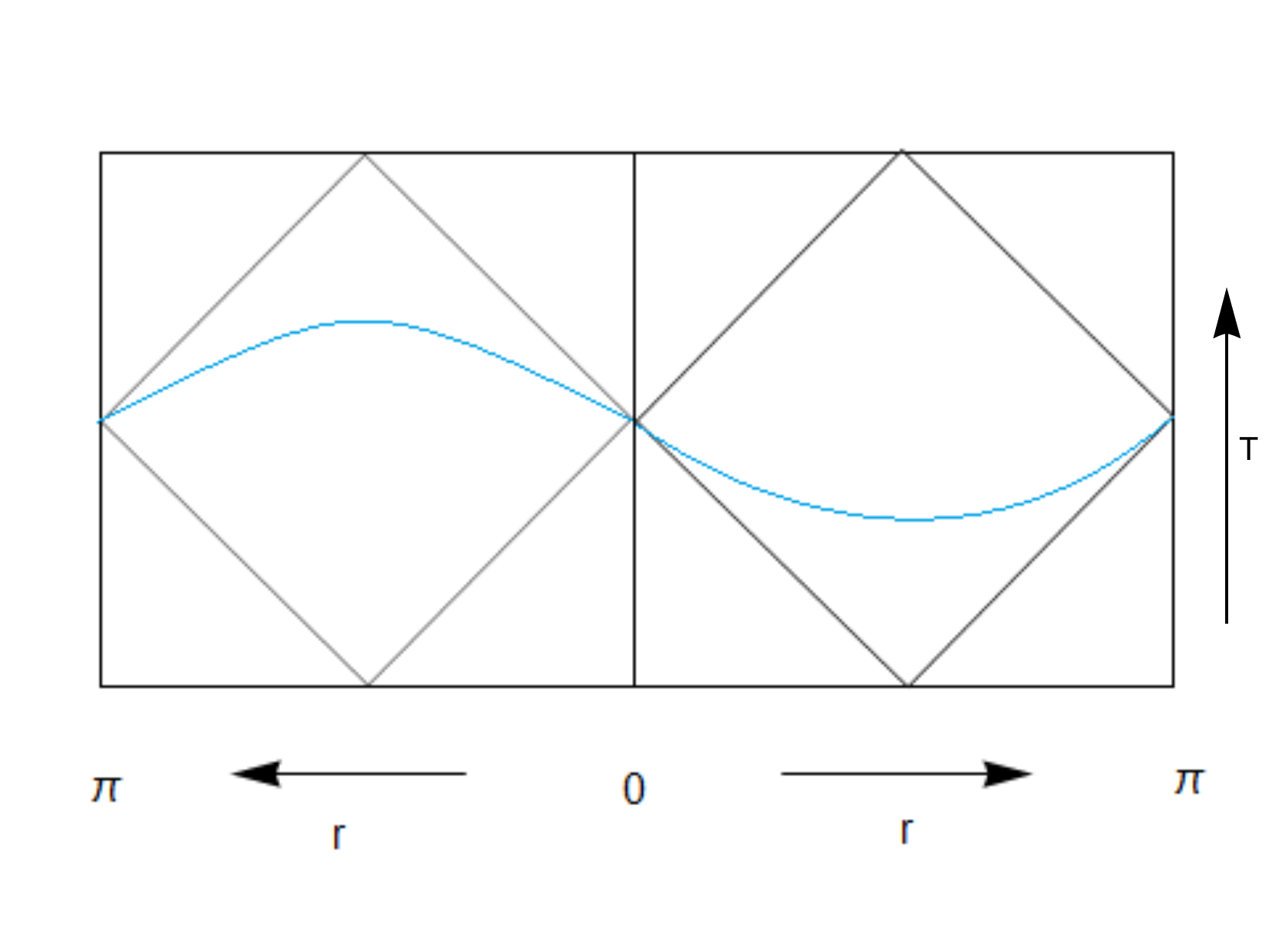}
        \caption{The Penrose Diagram of de-Sitter extended static patch sliced de-Sitter. To show the difference with de-Sitter global patch sliced de-Sitter, we draw it as two copies of the original diagram and the two edges on the left and right should be identified. Now each point represents a (D-2)-hemisphere. The left panel represents those with $\beta_{d}\in[0,\pi/2]$ and the right one represents those with $\beta_{d}\in[\pi/2,\pi]$. The blue slice represents a constant $t_{d}$ slice where the change of direction in global conformal time T can be seen from Equ.\ref{eq:transformation2}.}\label{pic:DS/dSextendedstatic}
\end{figure}

Instead of using global patch for those (D-1)-dimensional de-Sitter slices, we can use the extended static patch and we have the following coordinates parametrization
\begin{equation}
    \begin{split}
        X_{1}&=\cos{r}\\
        X_{0}&=\sin{r}\cos{\beta_{d}}\sinh t_{d}\\
        X_{2}&=\sin{r}\cos{\beta_{d}}\cosh t_{d}\\
        X_{k}&=\sin{r}\sin{\beta_{d}}\text{  }\Omega_{k}\label{equ:Ds/dSextendedstatic}
    \end{split}
\end{equation}
where $r,\beta_{d}\in[0,\pi]$, $k=3,\cdots,D$, $t_{d}$ is called the ``dS static time'' and we have the Penrose diagram Fig.\ref{pic:DS/dSextendedstatic} (for details see the caption). To understand the Penrose diagram the following coordinate transformation will be important
\begin{equation}
\begin{split}
    \sinh{\tau_{D}}&=\sin{r}\cos{\beta_{d}}\sinh{t_{d}}=\tan T\\
    \sin{\beta_{d}}&=\cosh{\tau_{d}}\sin{\chi_{d}} \label{eq:transformation2}
    \end{split}
\end{equation}
and using Equ.\ref{eq:transformation1} we have
\begin{equation}
   \sin{\beta_{d}}=\sqrt{\frac{\sin^{2}{\chi}}{\cos^{2}{T}-\cos^{2}{\chi}}}\sin{\chi_{d}}.
\end{equation}
This tells us that the hemisphere at each point in Fig.\ref{pic:DS/dSextendedstatic} is only part of the hemisphere at each point if we use the dS global patch to do DS/dS slicing. The reason that it is still a hemisphere is that at each point, when we use dS global patch, after we cut out and drop a (D-2)-annulus of that hemisphere in $\chi_{d}$ coordinate we have a local deformation, which can be understood as the factor $\sqrt{\frac{\sin^{2}{\chi}}{\cos^{2}{T}-\cos^{2}{\chi}}}$, and this scales the spherical cap back to a hemisphere. It is interesting to notice that antipodal points are still antipodal after this deformation.
\subsection{The DS/dS correspondence}

The DS/dS correspondence is based on the observation in \cite{Karch:2003em} that de-Sitter space can be engineered as the gluing of two ultra-violetly cut off anti-de Sitter spaces along an UV brane on which a lower dimensional residual graviton mode is localized due to the Randall-Sundrum mechanism \cite{Randall:1999vf}. Motivated by this observation and the AdS/CFT correspondence \cite{Maldacena:1997re,Gubser:1998bc,Witten:1998qj,Aharony:1999ti}, a holographic dual of dS$_{D}$ quantum gravity is proposed in \cite{Alishahiha:2004md} as a system of two conformal field theories, with a UV cutoff, living on the UV brane and they are coupled to each other by the localized graviton mode. The geometry of the UV brane is dS$_{D-1}$. 

The most straight forward way to understand this proposal is by the observation that both dS$_{D}$ and AdS$_{D}$ can be written as a foliation of dS$_{D-1}$ slices
\begin{equation}
    ds^{2}_{\text{(A)dS$_{D}$}}=dr^{2}+\text{sin(h)}^{2}(\frac{r}{L})\text{ }(-d\tau_{d}^{2}+\cosh^{2}(\frac{\tau_{d}}{L})d\Omega_{D-2}^{2})\label{equ:motivation}
\end{equation}
where for dS we used the DS/dS$_{global}$ coordinate \ref{equ:DS/dsglobal} and for AdS we used the so called Rindler patch. Here $r$ is the radial coordinate in the case of AdS and it is an energy scale in the context of the AdS/CFT correspondence where $r=0$ is the infra-red (IR) limit and $r=\infty$ is the UV regime. As the story goes, the AdS/CFT correspondence can be understood as defining the theory of quantum gravity in asymptotically AdS spaces. As we can see from the foliation Equ.\ref{equ:motivation} that the theory of quantum gravity in dS$_{D}$ and AdS$_{D}$ have the same IR physics which is the IR limit of the CFT living on the UV boundary of AdS$_{D}$. Then combined with the observation in \cite{Karch:2003em} we can see that the field theory dual of the quantum gravitational theory of dS$_{D}$ is a system of two CFTs living on the UV brane or the central slice ($r=\frac{\pi}{2}L$) coupled to each other via gravity. Recent development of the details of this DS/dS correspondence at an engineering level includes \cite{Gorbenko:2018oov,Lewkowycz:2019xse}.

\section{The Exact Non-local Entanglement Structure}
\label{sec:ENES}
DS/dS correspondence \cite{Alishahiha:2004md,Karch:2003em} tells us that quantum gravitational theory in a D-dimensional de-Sitter space (DS) is dual to a field theory system living on the central (D-1)-dimensional de-Sitter slice (dS), with residual graviton, $r=\frac{\pi}{2}L$, in the DS/dS patches Equ.\ref{equ:DS/dsglobal} or Equ.\ref{equ:Ds/dSextendedstatic}. Built on the observation that the entangling surfaces are great spheres in \cite{Geng:2019bnn}, we used the so-called surface/state correspondence \cite{Miyaji:2015yva} in \cite{Geng:2019ruz} to show that for pure DS space at zero-time slice $\tau_{D}=\tau_{d}=t_{d}=0$ the dual dS system has an exact non-local entanglement structure that antipodal points are maximally entangled\footnote{Other works addressing the non-locality of de-Sitter entanglement include \cite{Nomura:2019qps,Nomura:2017fyh}.}. However, there we used the DS/dS$_{\text{global}}$ patch Equ.\ref{equ:DS/dsglobal} where the dS global time $\tau_{d}$ is not an orbit of a time-like Killing vector field. This means that to study entangling surfaces beyond the zero-time slice we have to use the covariant holographic entanglement entropy proposal \cite{Hubeny:2007xt} and this is studied in \cite{Dong:2018cuv}. Let's firstly review the surface/state correspondence and what we did in \cite{Geng:2019ruz} and then we provide a way out of using the complicated covariant entropy proposal to extend the non-local entanglement structure beyond the $\tau_{d}=0$ slice. 

\subsection{Review of Previous Works}\label{sec:review}
In our previous works \cite{Geng:2019ruz,Geng:2019bnn} we studied de-Sitter holography on the zero time slice $\tau_{d}=0$. We noticed that the spatial geometry is spherical and according to the DS/dS correspondence the field theory dual is living on the equator. Therefore for the field theory subsystem half as large as the whole there are an infinite number of entangling surfaces with equal area and they are just half great spheres.

The surface/state correspondence\footnote{This proposal is generalized to a covariant version by \cite{Nomura:2018kji,Nomura:2016ikr}. } states that for any D-dimensional gravitational spacetime there is a correspondence between states in its Hilbert space and (D-2)-dimensional convex surfaces. Here convex means that if a surface $\Sigma$ is closed then extremal surfaces $\Gamma$ ending on any of its co-dimension one submanifold are totally included in the region surrounded by $\Sigma$ and if $\Sigma$ is an open surface it should be part of a closed convex surface $\Sigma_{c}$. This is supposed to be a generalized holographic principle \cite{tHooft:1993dmi,Susskind:1994vu} motivated by the formula relating entanglement entropy and the area of bulk extremal surfaces \cite{Ryu:2006bv,Ryu:2006ef}. On a practical level, it works as following. A closed convex surface $\Sigma$ corresponds to a pure state $\ket{\psi(\Sigma)}$ if it is topologically trivial (meaning that there is no spacetime singularity surrounded by $\Sigma$) and an open convex surface $\Sigma$ is dual to a density matrix $\rho(\Sigma)$ which is obtained by tracing out its complement $\tilde{\Sigma}$ on the corresponding closed convex surface $\Sigma_{c}$ from the pure density matrix $\ket{\psi(\Sigma_{c})}\bra{\psi(\Sigma_{c})}$ and the entanglement entropy between $\Sigma$ and $\tilde{\Sigma}$ is given by the RT formula \cite{Ryu:2006bv,Ryu:2006ef} as a quarter of the area of an extremal surface (in Planck unit) $\Gamma_{\Sigma}$ homologous to $\Sigma$ and anchored on its boundary $\partial\Sigma$. Moreover, any closed convex surface homologous to $\Sigma$ corresponds to the same state $\ket{\psi(\Sigma)}$. Furthermore, this correspondence gives an interpretation of the area $A(\Sigma)$ of a convex surface $\Sigma$. The area measures the number of local degrees of freedom participating in the entanglement of the state $\rho(\Sigma)$ (or single body states participating in the entanglement of the many body state $\rho(\Sigma)$). This entanglement includes the entanglement among the dual degrees of freedom living on $\Sigma$ and with those degrees of freedom on its complement $\Sigma_{c}$. This says that if $\Sigma$ is an extremal surface then there is no entanglement among the dual degrees of freedom living on it. In other words we can write $\rho(\Sigma)=\otimes_{i}\rho(\Sigma_{i})$ where $\Sigma_{i}$ are small segments making up $\Sigma$. This is how much we need from the surface/state correspondence.

Now applying the surface/state correspondence to the $\tau_{d}=0$ slice. The spatial geometry is spherical and so all extremal surfaces are co-dimension one great spheres including the equator where the dual field theory is living. Then the surface/state correspondence tells us that the state of the dual field theory has no local entanglement at all and the only entanglement it has is the maximal entanglement between those antipodal degrees of freedom. This important observation resolves several puzzles about entanglement entropy, entanglement of purification and complexity in de-Sitter holography as we formulated and discussed in \cite{Geng:2019ruz}. 

\subsection{Extending the Exact Non-local Entanglement Structure to Global Patch}
Now we want to extend the previous analysis of entanglement structure beyond the zero-time $\tau_{d}=0$ slice. Obviously, if we use coordinate systems that the derivative along its time-direction is not a Killing vector field, we have to use the covariant holographic entanglement proposal \cite{Hubeny:2007xt} to analyse and identify the entangling surfaces. And in this procedure we will lose the geometric intuition that we reviewed in Sec.\ref{sec:review}.

\begin{figure}
    \centering
   \includegraphics[width=10.1cm]{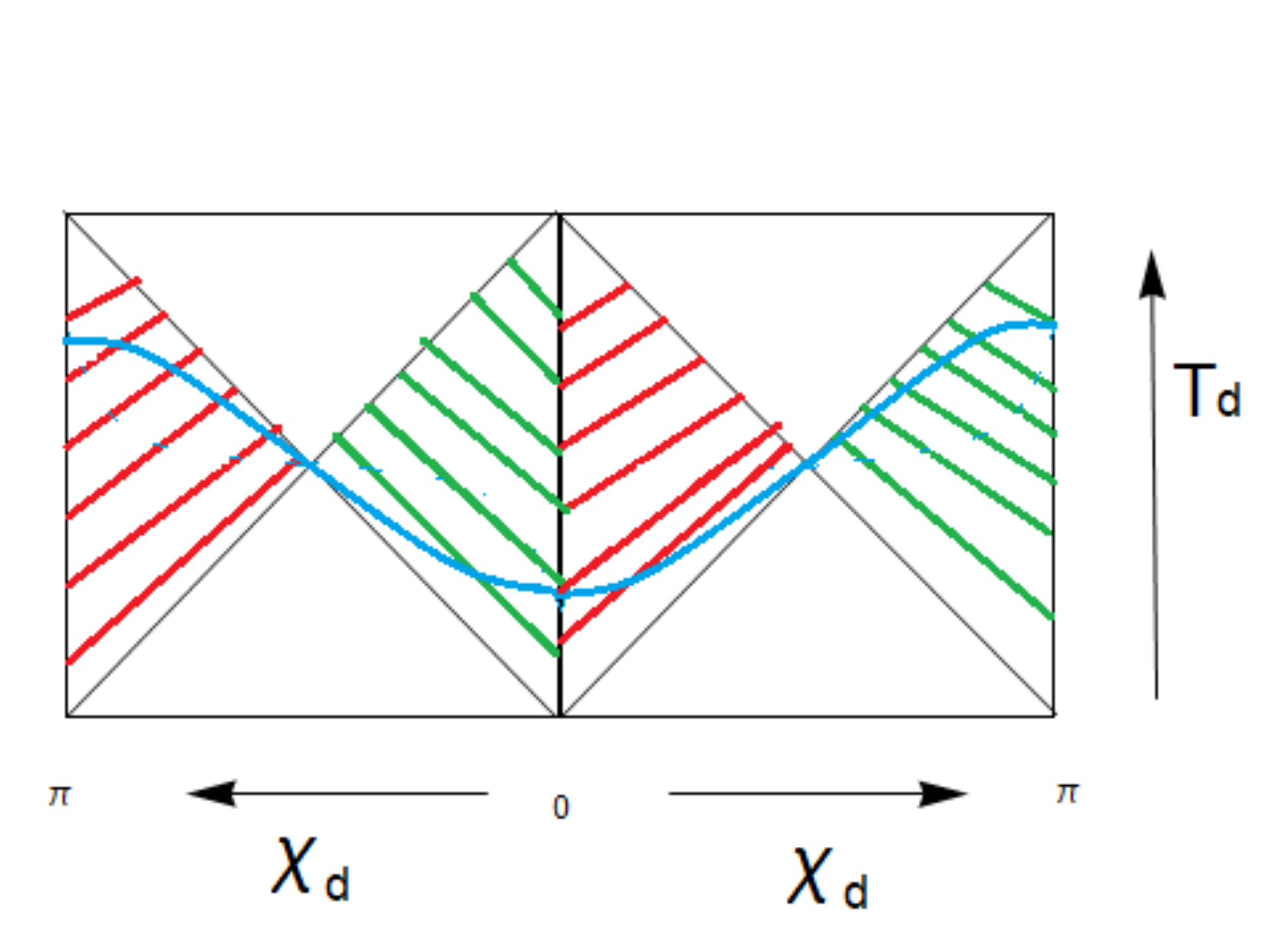}
        \caption{The Penrose Diagram of a (D-1)-dimensional de-Sitter (dS) where the field theory system is living on. A constant static time $t_{d}$ slice is shown in blue. Wedges with the same color are antipodally entangled at any constant-time slices. In order to show the non-local nature of the entanglement we draw two panels and each point of the diagram is a (D-3)-hemisphere.}\label{pic:antipodal2}
\end{figure}

Interestingly, we can analyse the entanglement structure beyond the zero-time slice by using the dS extended static patch to slice the D-dimensional de-Sitter (DS) Equ.\ref{equ:Ds/dSextendedstatic}. Here the dS static time $t_{d}$ is an orbit of a time-like Killing vector field in both dS and DS so there is no time dependence as long as we consider constant time slices. Again, the spatial geometry of the DS space is spherical so we have antipodal entanglement for each constant static time slice in dS (see Fig.\ref{pic:antipodal2}). Moreover, we can go one dimensional lower i.e. further slice dS using (D-2)-dimensional de-Sitter slices and use the DS/dS correspondence again (more precisely this time it should be dS/(d-1)S) and we see that in this lower dimensional system we again have antipodal entanglement. This a very important observation which tells us that the DS/dS correspondence \cite{Alishahiha:2004md,Karch:2003em} actually works in a way that a pure\footnote{Here ``pure" comes from our previous paper \cite{Geng:2019ruz} and it simply means that we consider the vacuum of the gravitional system so there is no excitation, for example a de-Sitter isometry-breaking solution of some quantum field, and the system is not deformed, for example by the $T\bar{T}$-type deformation \cite{Gorbenko:2018oov} whose effect is to put some spatial cutoff on the geometry.} D-dimensional de-Sitter space is dual to a pure (D-1)-dimensional de Sitter space (both of them are gravitational vacua with no spatial cutoff)! We call this observation the ``cascade of de-Sitter quantum gravity". 

\begin{figure}
    \centering
   \includegraphics[width=10.1cm]{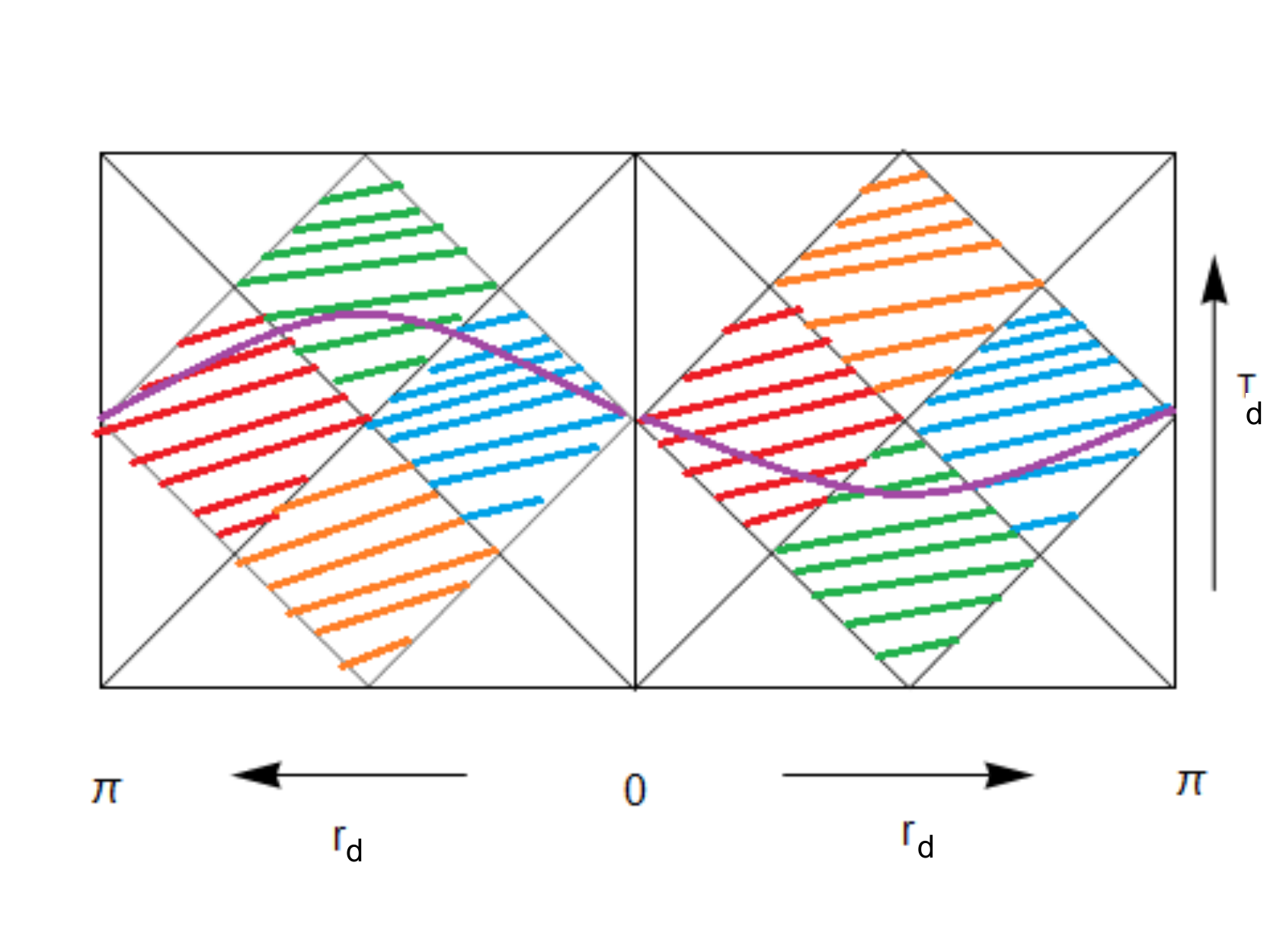}
        \caption{The entanglement structure motivated by the dS/(d-1)S$_{\text{extended static}}$. Patches with the same color are antipodally entangled on constant $t_{D-2}$ slices and each point on the diagram is a (D-3)-hemisphere.The purple curve is a constant $t_{D-2}$ slice.}\label{pic:antipodal1}
\end{figure}

\begin{figure}
    \centering
   \includegraphics[width=10.1cm]{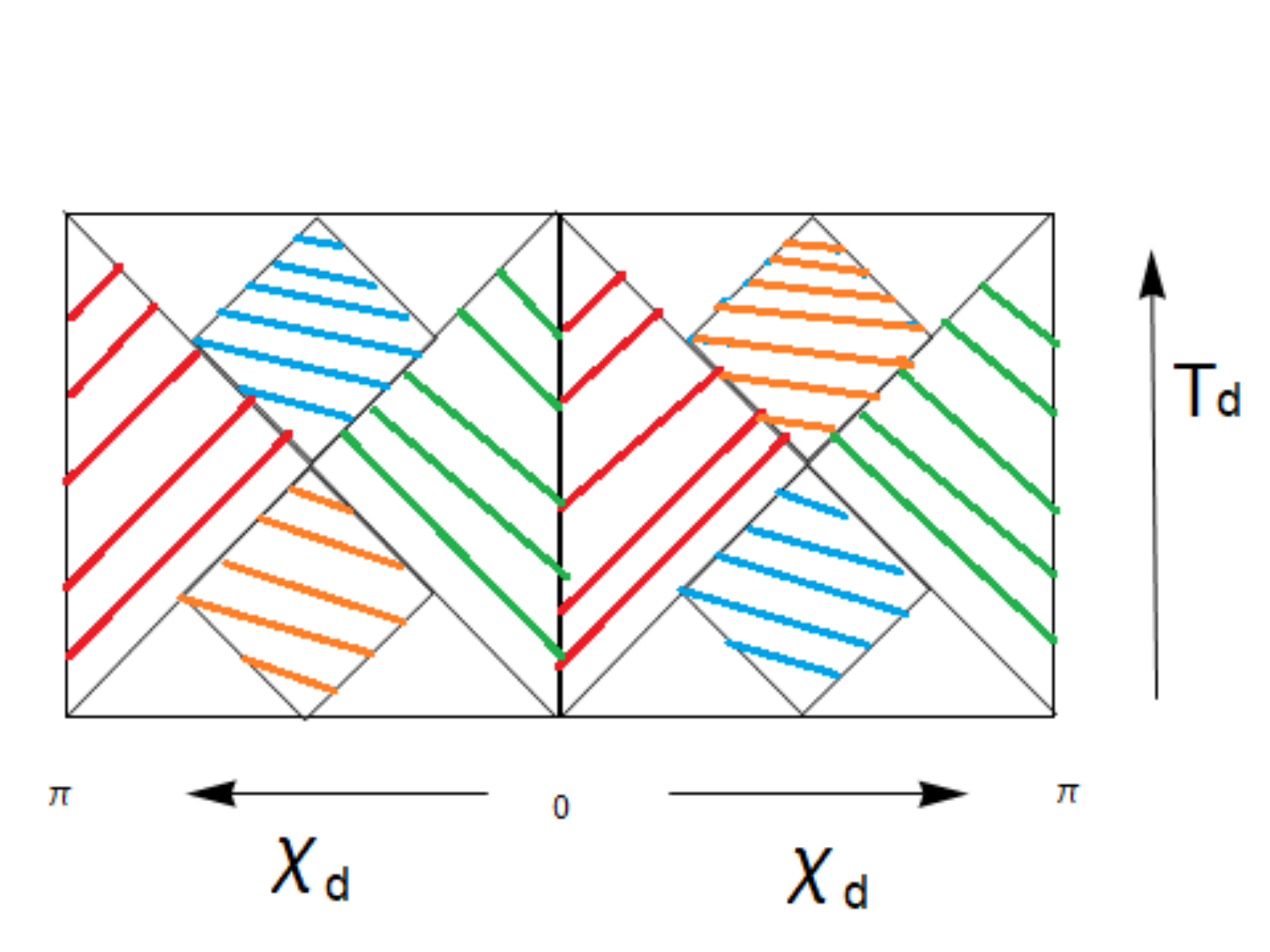}
        \caption{The ``maximally extended" entanglement structure. Each point on the diagram is a (D-3)-hemisphere. The diagram should now be understood as the global coordinate of a (D-1)-dimensional de-Sitter space.}\label{pic:antipodal3}
\end{figure}

Now the question is that can we extend this entanglement structure beyond the extended static patch? The answer is yes but we cannot extend it to the whole global patch. The way that we could proceed is inspired by our analysis of the ``cascade of de-Sitter quantum gravity" at the end of the previous paragraph. We can use the dS/(d-1)S$_{\text{extended static}}$ slicing for each dS slice of DS. In this coordinate system a time-like Killing vector field, whose orbit is the time coordinate, exits on the central diamonds of dS. Moreover, this is also a time-like Killing vector field in DS so do the same analysis as we did in DS/dS$_{\text{extended static}}$ this tells us that the antipodal entanglement structure on the dS extended static patch can be extended to those two central diamonds in Fig.\ref{pic:antipodal1} (again remember that we are talking about a (D-1)-dimensional de-Sitter so now each point is a (D-3)-hemisphere). Now if we recover the global coordinate for (d-1)S i.e. going to the dS/(d-1)S$_{\text{global}}$ patch, we will see that at each point of dS the entanglement structure only includes degrees of freedom living on a (D-3)-disk (or spherical cap) concentric to each (D-3)-hemisphere because of the coordinate transformation Equ.\ref{eq:transformation2}. Amazingly, using rotational symmetry we can adjust the part of each (D-3)-sphere showing up as those two antipodal (D-3)-disks and hence we can extend the antipodal entanglement structure from dS extended static patch to dS extended static patch union the dS/(d-1)S$_{\text{global}}$ patch! (See Fig.\ref{pic:antipodal3})

The take home message in this section is that DS/dS correspondence tells us some physics beyond the DS/dS patch (the central diamond). This nonlocal entanglement structure has important implications for the scrambling behavior of the dynamics of de-Sitter quantum gravity which will be studied in the next section.

\section{Shock Wave, Traversable Wormhole and Fast Srambling}\label{sec:fastscra}
The motivation of this section is from interesting works \cite{Shenker:2014cwa,Shenker:2013pqa,Roberts:2014isa} which studied chaos and scrambling of AdS BTZ black holes using holographic tools. A recent work trying to put forward those stories to de-Sitter space is \cite{Aalsma:2020aib}. The study in \cite{Aalsma:2020aib} focus on the static patch of de-Sitter. Even though it is lack of a well-defined holographic description, it makes the holographic study easy because people know how to describe black holes in de-Sitter space using the static patch and then, as it is done in \cite{Aalsma:2020aib}, we can use techniques developed for studies in AdS black holes \cite{Shenker:2013pqa,Shenker:2014cwa,Stanford:2014jda}. However, we will not follow that line of study because we want to work in the context of the DS/dS correspondence\cite{Alishahiha:2004md} which is a well-established holographic framework. The disadvantage in our approach is that we cannot do parallel studies as those already done in \cite{Aalsma:2020aib} because we don't know how to describe de-Sitter black holes in DS/dS and this prevents us from using those technologies developed before in the studies of AdS black holes. Nevertheless, the big advantage of our approach is conceptual because in the DS/dS correspondence the field system dual of de-Sitter gravity provides a solid context to consider chaos and fast scrambling. Since this is a rather long section we firstly summarize our logic and results. For those reader only caring about technical details, they can now jump to Sec.\ref{sec:twosided} starting from which we develop our own technologies which never appeared in the literature before.

In this section we will use the DS/dS correspondence to study chaos and scrambling in de-Sitter space. We will firstly do the maximal Kruskal extension of de-Sitter space by which we see that the two-sided geometry connected by a wormhole similar to that for an AdS black hole as in \cite{Shenker:2013pqa} appears. Then we will study scrambling property of the dynamics using the two-sided mutual information and a two-sided out-of-time-order correlator (OTOC) when there is a localized shock wave, which has the interpretation as a long time evolution of a locally propagating bulk excitation as that by Shenker and Stanford in \cite{Shenker:2013pqa}. We will see that the shock wave renders the wormhole to be traversable and the information scrambling time saturates the fast scrambler bound proposed by Sekino and Susskind in \cite{Susskind:2011ap}. And the late-time growing exponent of the two-sided OTOC behaves in an interesting way that it saturates the maximal chaotic bound proposed by Madalcena, Shenker and Stanford in \cite{Maldacena:2015waa} but with a negative sign compared to the usual maximal Lyapunov exponent. Suprisingly, this negative sign of the late-time growing exponent and the traversability of the wormhole has their origin in the exact non-local entanglement structure that we discovered in the previous section and it tells us that we can do teleportation using fast scramblers as long as the information is stored in a non-local entanglement structure.

Moreover, the fact that as time evolves a large amount a locally propagating excitation (particle) can be described as a bulk shock-wave (see \cite{Shenker:2013pqa}) is a nature of the Hamiltonian. However, how the bulk geometry responses to this shock wave should really be a property of the state \footnote{Here we should be careful because we have different time coordinate so accordingly we have different Hamiltonian. The essential observation is that the entanglement structure is invariant under the Hamiltonian associated to $\frac{\partial}{\partial t_{d}}$ and so we believe that the state $\ket{\psi(t_{d}=0)}$ is a zero energy eigenstate of this Hamiltonian. However, the Hamiltonian that sent the shock wave along the horizon should be the Hamiltonian associated with $\frac{\partial}{\partial\tau_{d}}$.}. From here we see that different microscopic entanglement structures really correspond to different macroscopic geometries which realizes the so-called ER=EPR proposal \cite{Maldacena:2013xja}. Here in de-Sitter we don't have local entanglement at all so the wormhole should be easier to be open such that we can transmit information between entangled boundary locations. However, for a BTZ black hole which is described by a thermal field double (TFD) state of two conformal field theories (CFTs) \cite{Maldacena:2001kr}, even though we have non-local entanglement between those two spacelikely separated CFTs, each energy eigenstate of the CFTs should have a strong local entanglement. As a result, for BTZ black holes there is still some local entanglement which makes the wormhole not able to be opened by bulk shock waves (as confirmed by \cite{Shenker:2013pqa}) but should be able to be opened by some non-local coupling between those two CFTs (as that studied by Gao, Jafferis and Wall in \cite{Gao:2016bin} and Maldacena and Qi in \cite{Maldacena:2018lmt}) because this non-local coupling enhances the non-local entanglement of that thermal field double state.

\subsection{The Two-sided Geometry}\label{sec:twosided}
\begin{figure}
    \centering
   \includegraphics[width=10.1cm]{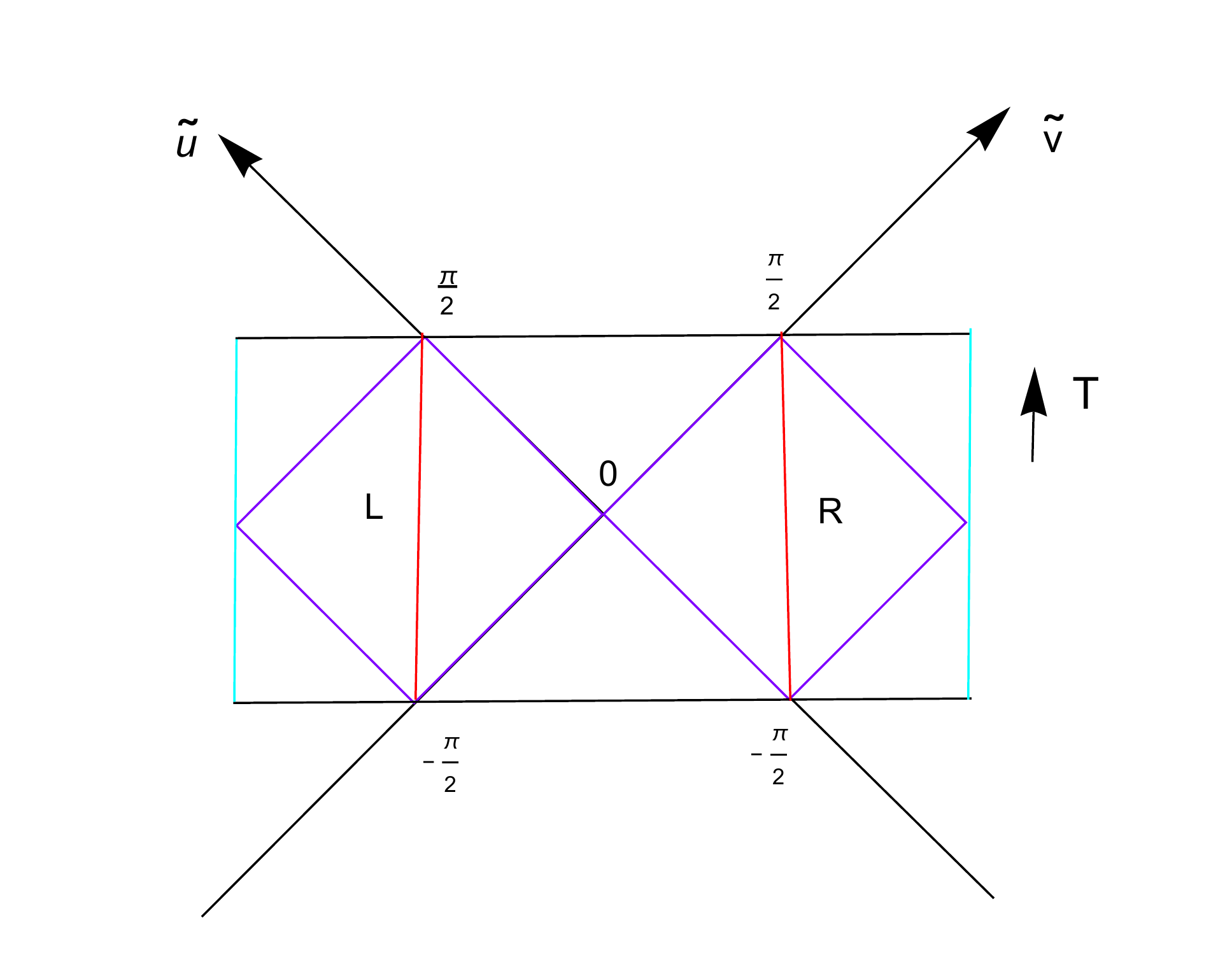}
        \caption{The Penrose diagram for the two-sided geometry. The two blue lines $\tilde{u}=\tilde{v}\pm\pi$ should be identified. The field theory system is living on the red slices L and R. The purple lines are horizons. The global conformal time T is going up.}\label{pic:twosided}
\end{figure}

To proceed we firstly formulate the DS/dS correspondence in terms as a two-sided bulk geometry such that on the bulk Penrose diagram the field theory system is living on two spacelikely separated edges. We use the DS/dS$_{\text{global}}$ patch Equ.\ref{equ:Ds/dSextendedstatic} which has the metric
\begin{equation}
    ds^{2}=dr^{2}+\sin^{2}{r}(-d\tau_{d}^{2}+\cosh^{2}{\tau_{d}}d\Omega_{D-2}^{2}).
\end{equation}
Then we following the normal procedure to find the Kruskal coordinates. We will keep $\chi_{d}$ to be a constant but it will have a $\pi$ jump as we go across $r=0,\pi$ which is a feature of the spherical geometry. In the tortoise coordinate we have
\begin{equation}
    ds^{2}=\sin^{2}{r}(dr_{*}^{2}-d\tau_{d}^{2}+\cosh^{2}{\tau_{d}}d\Omega_{D-2}^{2}),\text{   }r_{*}=\ln{\sqrt{\frac{1-\cos{r}}{1+\cos{r}}}}.
\end{equation}
Now we have the following null coordinates parametrizing geodesics going along $r-$direction,
\begin{equation}
    U=\tau_{d}+r_{*},\text{   }V=\tau_{d}-r_{*}\label{equ:kruskal1}
\end{equation}
which tells us that
\begin{equation}
    \cos{r}=\tanh{r_{*}}=\tanh{\frac{U-V}{2}},\text{   }\tau_{d}=\frac{U+V}{2}.\label{equ:Kruskal2}
\end{equation}
Now the metric is
\begin{equation}
    ds^{2}=-\frac{dUdV}{\cosh^{2}{\frac{U-V}{2}}}+\frac{\cosh^{2}{\frac{U+V}{2}}}{\cosh^{2}{\frac{U-V}{2}}}d\Omega_{D-2}^{2}.
\end{equation}
Then we have the following Kruskal coordinates and metric
\begin{equation}
   ds^{2}=-4\frac{dudv}{(1-uv)^{2}}+\frac{(u-v)^{2}}{(1-uv)^{2}}d\Omega_{D-2}^{2}, u=e^{U},v=-e^{-V}.\label{equ:Krushal3}
\end{equation}
In terms of the embedding coordinates in Equ.\ref{equ:DS/dsglobal}, we have
\begin{equation}
\begin{split}
    X_{1}&=-\frac{1+uv}{1-uv}=\cos{r}\\
    X_{0}&=\frac{u+v}{1-uv}=\sin{r}\sinh{\tau_{d}}\\
    X_{j}&=\frac{u-v}{1-uv}\Omega_{j}=\sin{r}\cosh{\tau_{d}}\Omega_{j}.
    \end{split}
\end{equation}
Now this coordinate system $(u,v)$ can be conformally embedded into $\mathbb{R}^{2}$ by the following transformation
\begin{equation}
    u=\tan{\tilde{u}},v=\tan{\tilde{v}}, \text{ where  } \tilde{u},\tilde{v}\in [-\pi,\pi]\label{equ:maximalKruskal}
\end{equation}
and the Penrose diagram is given in Fig.\ref{pic:twosided}. We should be careful when we look at this diagram because this diagram effectively extends $r$ from $[0,\pi]$ to $[0,2\pi]$. There are two ways to understand this and they give the same result. The first is a mathematical way where now each point on the diagram should be a hemisphere instead of a sphere because of this extension and the spherical geometry. The second way is a physical way that this two sided geometry represents a thermal field double state ($\ket{\text{TFD}}$) of the two field theory systems living on the red lines with temperature $T_{dS}=\frac{1}{2\pi}$. Hence these two systems are entangled to each other. But from our analysis of the non-local entanglement structure we know that each point on the field theory system is maximally entangled with its antipodal point. Then using the monogamy of entanglement we see that the two red lines must be antipodal partners of each other and therefore each point on the diagram must be a hemisphere. Furthermore, we should notice that, even though the global conformal time is going up, the dS global time $\tau_{d}$ is going oppositely inside the two diamonds.

\subsection{Shock Wave Geometry}\label{sec:shockwave}
\begin{figure}
    \centering
   \includegraphics[width=10.1cm]{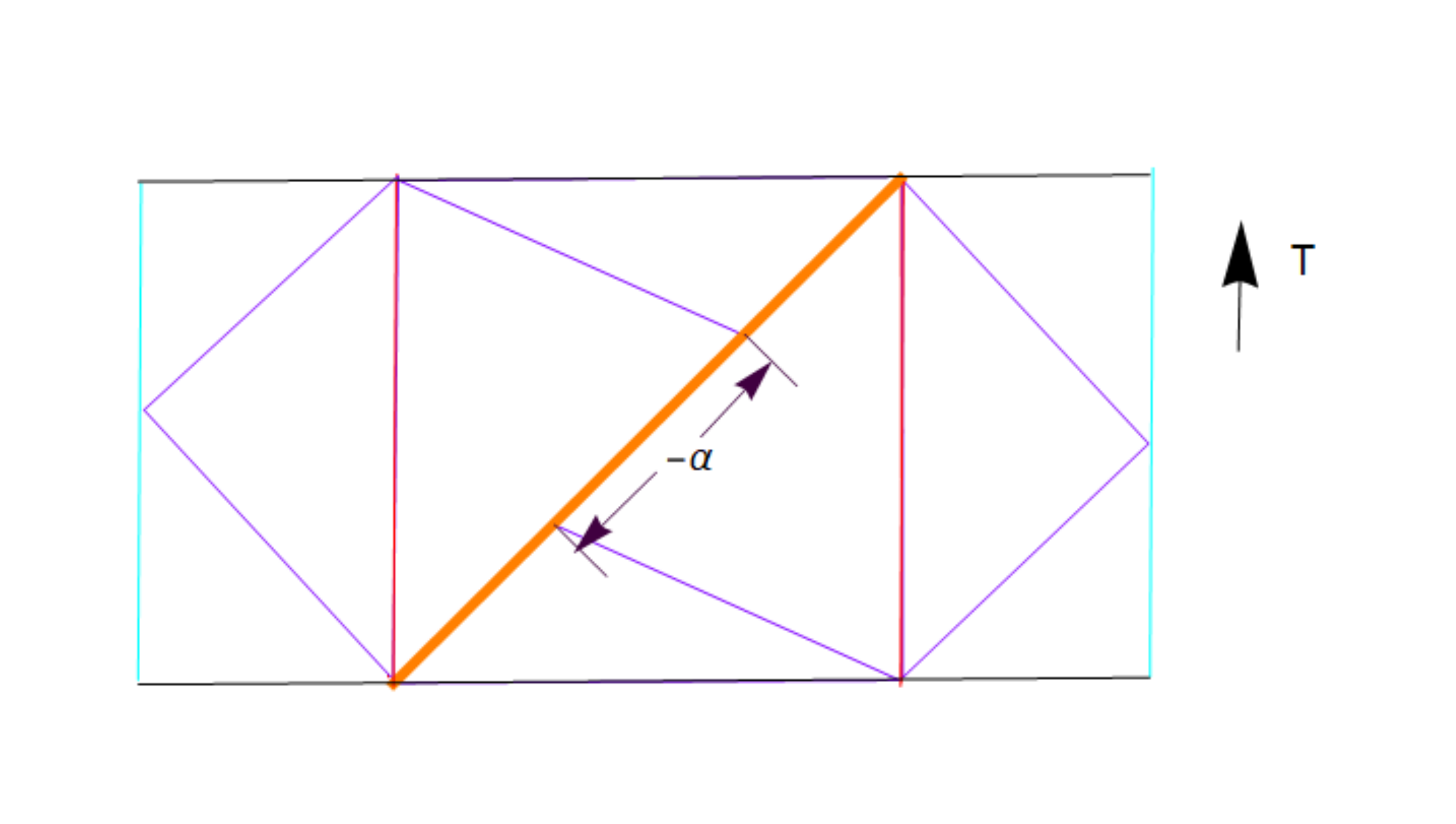}
        \caption{The Penrose diagram for the shock wave geometry Equ.\ref{equ:shockwave}. The orange line is the shock wave.}\label{pic:shockwave}
\end{figure}

For the sake of simplicity, we consider $DS_{3}$ (as the bulk). To probe the information scrambling, we drop a few quanta from the left field theory system (L in Fig.\ref{pic:twosided}) at time $\tau_{d}=\tau_{w} (<0)$ long in the past with energy $E$ along the radial direction and study how this information is scrambled among the system at the moment $\tau_{d}=0$. If we measure the energy of these quanta at $\tau_{d}=0$ slice in its local frame, we will see the proper energy
\begin{equation}
    E_{p}\sim \frac{E}{2}e^{-\tau_{w}}.
\end{equation}
Note that the quanta only go along the radial direction $r$ which means that $\frac{\partial}{\partial\tau_{d}}$ is an effective time-like Killing vector field (remember the metric Equ.\ref{equ:DS/dsglobal}) so the story of blue-shifted energy from the boundary observer to a bulk local observer goes the same as that in Schwartzschild black hole \cite{Wald:1984rg}. In the local bulk frame the perturbation has very high (blue shifted) energy which can be described by a shock wave going along a null trajectory $v=\text{const.}$ near the horizon\footnote{Intuitively, this can be seen by using the fact that as $\tau_{d}$ goes to the past $u$ is compressed and $v$ is stretched (see Equ.\ref{equ:kruskal1}, Equ.\ref{equ:Kruskal2} and Equ.\ref{equ:Krushal3}).}  \cite{Liu:2013iza,Roberts:2014isa}. In a mathematically manner, this can be seen as following. We assume that the backreaction of these quanta on the geometry is a shift of the $v$ coordinate when $u>0$ (the reason is that $u,v$ are perpendicular null directions and the only nonzero stress-energy tensor we would have is $T_{uu}$ and so comparing to pure de-Sitter the only changed component of the Ricci curvature is $R_{uu}$):
\begin{equation}
\begin{split}
    X_{1}&=-\frac{1+\big[v+\alpha\theta(u)\big]u}{1-\big[v+\alpha\theta(u)\big]u}\\
    X_{0}&=\frac{\big[v+\alpha\theta(u)\big]u}{1-\big[v+\alpha\theta(u)\big]u}\\
    X_{2}&=\frac{u-\big[v+\alpha\theta(u)\big]}{1-\big[v+\alpha\theta(u)\big]u}\cos{\phi}\\
    X_{3}&=\frac{u-\big[v+\alpha\theta(u)\big]}{1-\big[v+\alpha\theta(u)\big]u}\sin{\phi}
    \end{split}\label{equ:shockwave}
\end{equation}
where $\phi\in[0,\pi]$ since each point is a hemisphere. Then we can work out the metric in the embedding space formalism:
\begin{equation}
    ds^{2}=\frac{-4dudv-4\alpha\delta(u)du^{2}+(u-v-\alpha\theta(u))^{2}d\phi^{2}}{\Bigg[1-\big[v+\alpha\theta(u)\big]u\Bigg]^{2}}.
\end{equation}
Using Einstein's equation we find that the only non-zero component of the stress-energy tensor is
\begin{equation}
    T_{uu}=-\frac{\alpha}{2\pi G}\delta(u).
\end{equation}
Now the null energy condition tells us that $\alpha<0$ and from the relation between and the components of stress-energy tensor
\begin{equation}
    \int d^{2}x T^{00} |_{\tau_{d}=0}=E_{p}
\end{equation}
we see that
\begin{equation}
    -\alpha= \mathcal{O}(1) GE_{p}=\mathcal{O}(1)GE e^{\frac{-\tau_{w}}{L}},
\end{equation}
where we restored the $L$ dependence in this final expression. This tells us that the geometry responds in a way that the two field theory systems are now in causal contact see Fig.\ref{pic:shockwave}. This is consistent with the intuition that the non-local entanglement is very strong so we don't have to further enhance the non-local entanglement by introducing extra coupling between the two sides as that in the AdS BTZ black hole \cite{Gao:2016bin} to blow up the wormhole. More importantly, this realizes the idea that ER=EPR or geometry (and its response to local perturbations) is actually a macroscopic representation of entanglement \cite{Maldacena:2013xja}. 

\subsection{Geodesics}
In de-Sitter space the geodesic distance $D(x,y)$ between two points $x$ and $y$ is most easily dealt with using their embedding space coordinates $X$ and $Y$ (see for example \cite{Strominger:1996sh}) and for DS$_{3}$ it is given by the $SO(1,3)$ invariant:
\begin{equation}
    \cos{D(x,y)}=-X_{0}Y_{0}+X_{1}Y_{1}+X_{2}Y_{2}+X_{3}Y_{3}.
\end{equation}
The geodesics connecting two points at the boundaries L and R separately with dS global times $\tau_{L},\tau_{R}$ (both positive or both negative) and spatially antipodal to each other will go across the shock at a point with $u=0$ and some value of $v$. However, for L this point has $u=0^{+}$ and for R this point has $u=0^{-}$. These two distances are
\begin{equation}
    \begin{split}
        \cos{d_{L}}&=-(v+\alpha)\cosh{\tau_{L}}\\
        \cos{d_{R}}&=v\cosh{\tau_{R}}
    \end{split}\label{eq:gd}
\end{equation}
where we emphasize that we have to be careful about the angular coordinate $\phi$. If the angular coordiante is $\phi$ for the L point then for its antipodal point on R the angular coordinate should be $\pi+\phi$ and this $\pi$ shift accounts for the minus sign in Equ.\ref{eq:gd}. Now extremizing $d_{L}+d_{R}$ over $v$, we have the total geodesic distance between them
\begin{equation}
    d=2\cos^{-1}{\frac{-\alpha \cosh{\tau_{L}}\cosh{\tau_{R}}}{\cosh{\tau_{L}}+\cosh{\tau_{R}}}}
\end{equation}
where we notice that the null energy condition tells us that $\alpha<0$ and so this is smaller than their angular separation $\pi$ which is the distance of another geodesics connecting them. From here we can see that those two boundary points are timelikely separated ($d$ imaginary) if $-\alpha$ is large enough:
\begin{equation}
    -\alpha>\frac{1}{\cosh{\tau_{L}}}+\frac{1}{\cosh{\tau_{R}}}. 
\end{equation}
Also for a given value of $\alpha$ if $\tau_{L}$ and $\tau_{R}$ are large enough then the two points are also time-likely separated which is consistent with Fig.\ref{pic:shockwave}.

The geodesics distance between two equal-time points on the same side with angular separation $\phi$ (remember $\phi<\pi$) is given by
\begin{equation}
    d=\cos^{-1}(1-2\cosh^{2}{\tau}\sin^{2}{\frac{\phi}{2}}).
\end{equation}
At late time, this distance will be imaginary. This was also noticed in \cite{Dong:2018cuv} and following their discussion we rewrite this geodesic distance $d$ as
\begin{equation}
    d=2\tan^{-1}{\sqrt{\frac{\sin^{2}{\frac{\phi}{2}}\cosh^{2}\tau_{L}}{1-\sin^{2}{\frac{\phi}{2}}\cosh^{2}{\tau_{L}}}}}
\end{equation}
and at late time we take the limit $\sin^{2}{\frac{\phi}{2}}\cosh^{2}\tau_{L}\rightarrow1$ which means $d\rightarrow\pi$. Moreover, it is interesting to notice that if we use the DS/dS$_{\text{extended static}}$ patch Equ.\ref{equ:Ds/dSextendedstatic} then there is no such pathology because there we have a time like Killing vector field but there we cannot write it in terms of a two-sided geometry as we did in this section.
\subsection{Mutual Information Measures the Traversability of the Wormhole}
In this section, we will study the two-sided mutual information using holographic proposals \cite{Ryu:2006bv,Ryu:2006ef} and use it to probe information scrambling. We study two subsystems $A$ and $B$ at zero time with equal size $\phi$ and are antipodally located on the two sides. The mutual information between them is defined by
\begin{equation}
    I(A:B)=S_{A}+S_{B}-S_{A\cup B}
\end{equation}
where $S$ is the entanglement entropy. As the name suggests the mutual information measures the amount information about A or B that is encoded in the other. It is zero when there is no entanglement between A and B, i.e. $S_{A\cup B}=S_{A}+S_{B}$, because information in encoded in the entanglement. As we notice in \cite{Geng:2019ruz} that the RT proposal of picking up the smallest entangling surface works for connected intervals but not for disconnected intervals and when there is no shock wave $S_{A\cup B}$ should be zero. Hence for $S_{A\cup B}$, when there is shock wave, one will firstly use the RT proposal and then subtract its value when $\alpha=0$. This tells us that
\begin{equation}
\begin{split}
    S_{A}&=S_{B}=\frac{1}{2G}\Bigg[\tan^{-1}\sqrt{\frac{\sin^{2}{\frac{\phi}{2}}\cosh^{2}{\tau_{L}}}{1-\sin^{2}{\frac{\phi}{2}}\cosh^{2}{\tau_{L}}}}\Bigg]_{\tau_{L}=0}=\frac{\phi}{4G}\\
    S_{A\cup B}&=\frac{1}{2G}\big[2\cos^{-1}(-\frac{\alpha}{2})-\pi\big],
    \end{split}
\end{equation}
so the mutual information is
\begin{equation}
    I(A:B)=\frac{1}{2G}\big[\phi+\pi-2\cos^{-1}(\frac{-\alpha}{2})\big].
\end{equation}
This tells us that the mutual information grows as we increase $-\alpha$ which is roughly the size of the traversable wormhole\footnote{Note that we have the late time issue we discussed in the previous subsection and taking the strategy as there the maximal mutual information is $I(A:B)=\frac{\phi+\pi}{2G}$ which gives a bound on the size of the traversable wormhole or the amount of information we can teleport between the two sides. } which also measures the amount of quanta we drop in at the early time $\tau_{w}$.

It might be a bit puzzling that we know when there is no shock wave, by the exact non-local entanglement structure, the mutual information $I(A:B)$ takes it maximal value (because in that case $S_{A\cup B}=0$) then how could we further increase it or how could $S_{A\cup B}$ be smaller than zero? For the second question we don't have a very persuasive answer beyond an example in our mind that an regularization can render a seemingly positive series to negative- the zeta function regularization in string theory. But for the first question the answer is by noticing that for small $\alpha$ the amount of that further increase is proportional to $\alpha$ which is the amount of quanta we further added to the system at early time. And importantly this observation tells that the exact non-local entanglement structure is stable against local perturbations (the mutual information is a power series of $\alpha$ so there is no non-perturbative effects). In other words, the quanta we dropped in at early time will be scrambled all over the system such that they are also antipodally entangled.

\subsection{The Scrambling Time- de-Sitter Is a Fast Scrambler}
Using the fact that $-\alpha\propto E_{p}$ and restoring the dependence on the curvature scale $L$ we can see that the mutual information saturates the maximal bound when
\begin{equation}
-\tau_{w}=\tau_{*}=L\log(\frac{2}{GE}\mathcal{O}(1)).
\end{equation}
And using the fact that the de-Sitter temperature and entropy are
\begin{equation}
    T=\frac{1}{2\pi L}, S=\frac{\pi L}{2G},
\end{equation}
we can see that for the energy $E$ with its smallest possible value $E\sim T$ (note that in DS/dS the de-Sitter temperatures for DS and dS are the same so this approximation has a nice interpretation in the dS description)
\begin{equation}
    \tau_{*}\sim\frac{1}{2\pi T}\log(S).
\end{equation}
Interestingly, to be mathematically precise we have used the $\sim$ symbol but now for this scrambling time we can replace $\sim$ by $=$ because for a holographic field theory the entropy is large so we can ignore that unknown $\mathcal{O}(1)$ factor. This is the time that the system scrambles one thermal bit of information from one side to two sides. Moreover, the original non-local entanglement structure is stable against the scrambling so the scrambling did not destroy the mutual information as that for BTZ black hole \cite{Shenker:2013pqa}! This tells us that fast scrambling systems can be used to efficiently teleport if the information is encoded carefully in a non-local entanglement structure.
\subsection{Correlation Functions}
We can see the same physics as the two-sided mutual information told us using two-sided correlators. For a boundary operator dual to a heavy bulk scalar field, we can use the first quantization picture in the bulk to get the two-point function 
\begin{equation}
    \langle\phi_{L}(x)\phi_{R}(y)\rangle\sim e^{-md_{bulk}(x,y)}.\label{equ:2pt}
\end{equation}
For two antipodal points at zero time, with the L dependence restored, we have
\begin{equation}
    \langle\phi_{L}\phi_{R}\rangle\sim e^{-2m\cos^{-1}(-\frac{\alpha}{2L})}=e^{-2m\cos^{-1}(e^{-\frac{-\tau_{*}-\tau_{w}}{L}})}.
\end{equation}
Again it is unaffected by the perturbation until $-\tau_{w}\sim \tau_{*}$ when it stars to super-exponentially grow. However, when $-\tau_{w}\ll \tau_{*}$ or $-\alpha\ll L$ we have the following expansion
\begin{equation}
    \langle\phi_{L}\phi_{R}\rangle\sim e^{-\pi m}(1-\pi \frac{\alpha}{L}+\mathcal{O}((\frac{\alpha}{L})^{2}))
\end{equation}
which, to the leading order in $\alpha$, in terms of $\tau_{w}$ reads
\begin{equation}
    \langle\phi_{L}\phi_{R}\rangle\sim e^{-\pi m}(1+2\pi e^{-\frac{\tau_{*}}{L}}e^{-\frac{\tau_{w}}{L}}).
\end{equation}
We see that it exponentially increases with $-\tau_{w}$ with the exponent saturating the maximal chaos bound \cite{Maldacena:2015waa}
\begin{equation}
    \lambda_{L}=\frac{1}{L}=\frac{1}{2\pi T}.
\end{equation}

Another interesting interpretation of this simple calculation is to rewrite the two-point function that we computed as a two-sided out-of-time-ordered correlator (OTOC) and relate it to quantum chaos \cite{Shenker:2013pqa,Maldacena:2015waa}. Suppose that the boundary operator creating the shock wave is $W(\tau_{w})$, the boundary field theory state dual to pure de-Sitter is $\ket{\psi}$ and the boundary operator dual to the bulk field $\phi(x)$ is $V(0)$ (which should be Hermitian as $\phi(x)$ is Hermitian and 0 means zero time). Then the two-point function can be rewritten as
\begin{equation}
    \langle\phi_{L}(x)\phi_{R}(y)\rangle=\frac{\bra{\psi}W^{\dagger}(\tau_{w})V^{\dagger}_{L}(0)V_{R}(0)W(\tau_{w})\ket{\psi}}{\bra{\psi}W^{\dagger}(\tau_{w})W(\tau_{w})\ket{\psi}}
\end{equation}
where the spatial supports of $V_{L}(0)$ and $V_{R}(0)$ are antipodal to each other. This correlator is out-of-time-ordered. We emphasize that the fact that, as we have discussed in Sec.\ref{sec:shockwave}, the operator $W(\tau_{w})$ creates a shock wave is because it is inserted at late time ($\frac{1}{2\pi T}=\frac{1}{L}\ll-\tau_{w}$). And as opposite to the usual story of fast scramblers that at late time but before scrambling ($\frac{1}{2\pi T}\ll -\tau_{w}\ll \tau_{*}$) the local OTOCs are exponentially decreasing with time with the maximal Lyapunov exponent, here this two-sided OTOC is exponentially growing.

\subsection{A Cartoon Interpretation}
\begin{figure}
    \centering
   \includegraphics[width=10.1cm]{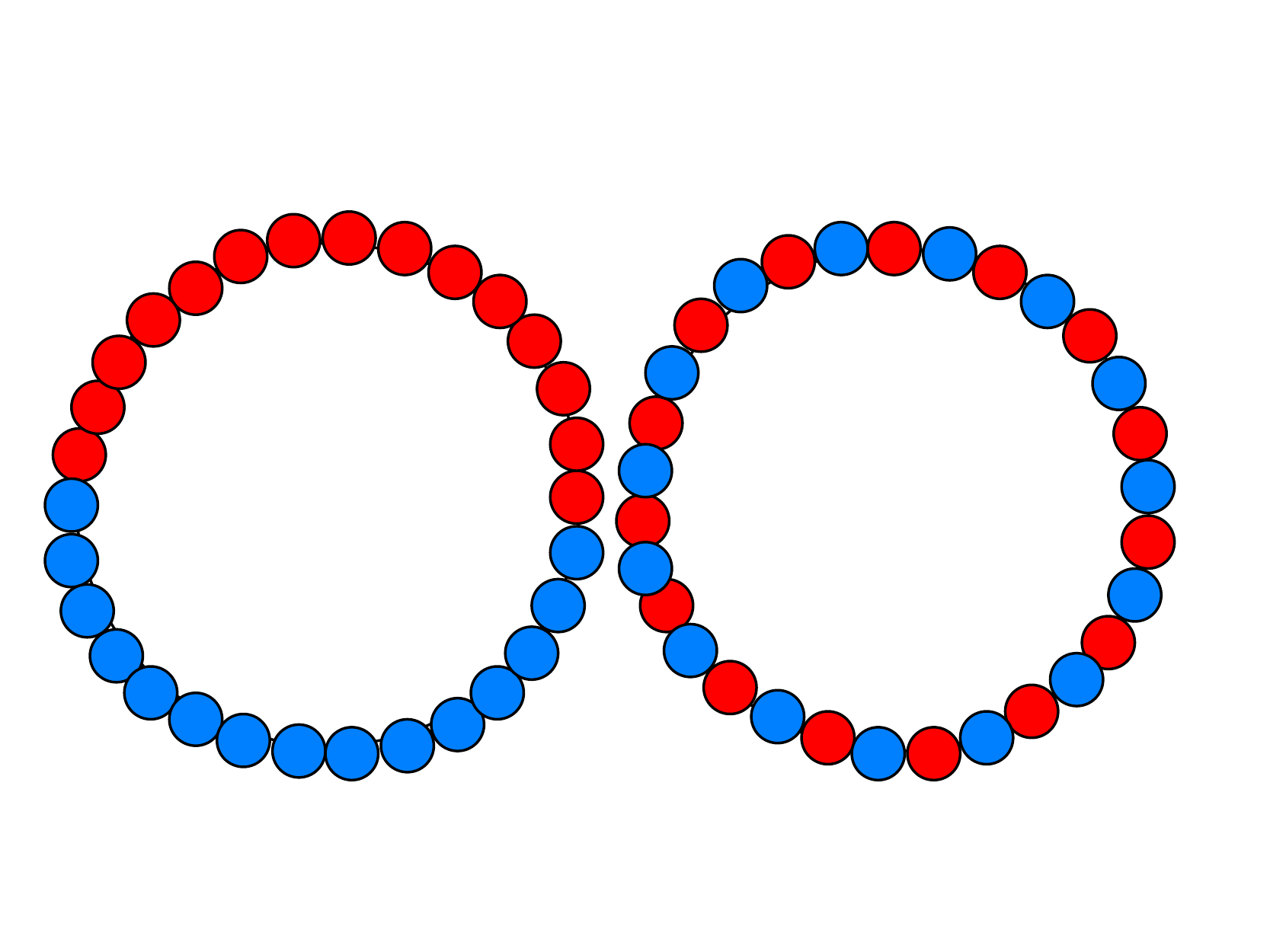}
        \caption{The left panel represents the state before scrambling and the right one represents a typical state after scrambling. Each antipodal pair has a blue ball and a red ball and the antipodal red and blue balls are maximally entangled. }\label{pic:scrambling}
\end{figure}

In Fig.\ref{pic:scrambling} we show the entanglement structure before and after scrambling. We can see that if the information is only encoded in this non-local entanglement structure then there is no information scrambling because the entanglement structure is antipodal before and after the scrambling . However the information encoded in a locally entangled pair will be scrambled (for example initially we insert a maximally entangled qubits locally at some point then after scrambling this piece of information is scrambled everywhere and typically the two qubits are antipodal to each other). This provides an interpretation for both the traversability of the wormhole and the growing of the two-sided OTOC or the negativity of its ``Lyapunov exponent".

\subsection{Teleportation Using Fast Scramblers}
From our study and the earlier studies in AdS, a traversable wormhole means the ability to do quantum teleportation \cite{Susskind:2017nto,Brown:2019hmk}. In AdS black hole if there is no coupling between the two boundaries then they cannot do classical communication\footnote{Classical communication is an essential element in quantum teleportation \cite{nielsen00}.} and hence the wormhole is not traversable. On the contrary, if we could do classical communication by either introducing proper coupling between the two boundary theories in AdS black hole or we use the dS system then, together with the fact that we have quantum entanglement between the two sides, then we are able to do quantum teleportation. Here it is worth to be emphasized that we can use the dS system because in this case the systems living on two sides of Fig.\ref{pic:twosided} are actually the same system as we have already discussed at the end of Sec.\ref{sec:twosided} using the non-local entanglement structure and the monogamy of entanglement.

\section{Conclusions and Future Remarks}\label{sec:final}
In this work, we extend the non-local entanglement structure in \cite{Geng:2019ruz} beyond the zero-time slice based on which we push forward the interesting stories of faster scrambling \cite{Shenker:2014cwa,Shenker:2013pqa} and traversable wormhole \cite{Gao:2016bin} in AdS BTZ black hole to dS space and we find interesting behaviors in our context very different from those in AdS black hole. As that in BTZ black hole using the AdS/CFT correspondence, our study has a well-defined holographic framework-the so-called DS/dS correspondence \cite{Alishahiha:2004md,Karch:2003em}. Moreover, we interpret our studies as providing an example that a specific microscopic entanglement structure is equivalent to a corresponding macroscopic geometry and this is a realization of the ER=EPR proposal \cite{Maldacena:2015waa}. This work inspires and paves the way for future studies of dS quantum gravity and more properties of its holographic dual, for example the time evolution of complexity and the two-sided entanglement entropy as those in AdS black hole \cite{Susskind:2014rva,Stanford:2014jda,Hartman:2013qma}.
\section*{Acknowledgement}
I appreciate useful discussions with Tarek Anous, Jing-Yuan Chen, Wenjie Ji, Andreas Karch, Tsung-Cheng Peter Lu, Alexander Maloney, Hao-Yu Sun, Gerard 't Hooft, Cenke Xu and Laurence Yaffe. This work was supported in part by a grant from the Simons Foundation (651440, AK). I am very grateful to my parents and recommenders.

\bibliographystyle{JHEP}
\bibliography{LREDS}
\end{document}